# An Analysis on Interactions among Secondary User and Unknown Jammer in Cognitive Radio Systems by Fictitious Play

Ehsan Meamari  Khadijeh Afhamisisi  Hadi Shahriar Shahhoseini

School of Electrical Engineering, Iran University of Science and Technology, Iran

ehsanmeamari@gmail.com, afhami@iust.ac.ir, hshsh@iust.ac.ir

*Abstract*—With the advancement of communication, the spectrum shortage problem becomes a serious problem for future generations. The cognitive radio technology is proposed to address this concern. In cognitive radio networks, the secondary users can access spectrum that allocated to the primary users without interference to the operation of primary users. Using cognitive radio network raises security issues such as jamming attack. A straightforward strategy to counter the jamming attack is to switch other bands. Finding the best strategy for switching is complicated when the malicious user is unknown to the primary users. This paper uses fictitious game for analysis the defense against such an unknown jammer.

## I. INTRODUCTION

The cognitive radio network (CRN) is introduced for next generation networks because of finite nature of the radio spectrum [1]. In CRN, secondary users can access to the frequency bands without interfering with primary users [2-3]. CRN as a new introduced technology raises security issues [4], such as jamming attack, which are needed to be addressed. To counter jamming attack, it is recommended to the secondary users to switch to other vacant bands when is jammed by the malicious users [5].

The optimal strategies against jamming attacks using Markov Decision Process (MDP) is proposed by Wu, Wang and Liu in [6]. Meamari et al. proposed a game theoretical method to analyze a coordinated jamming attack [7]. Tan, Sengupta and Subbalakshmi suggested a model to find the optimal bands of malicious users for switching [8]. Then, they utilized Markov model to demonstrate that the malicious users are more efficient against coordinated jamming attack than non-coordinated jamming attack.

The fictitious play can be used to analyze the attack when there isn't complete information about their rival [9]. the fictitious play is used to calculate the optimal strategies for different players of attack by Nguyen, Alpcan and Basar in [10].

We considered a simple model of CRN in which there is only one secondary and one malicious user exist there. The secondary user transmits its information to proposed network while the malicious user wants to attack by jamming. Afterwards, the optimal strategies are calculated for both secondary and malicious users in an incomplete information situation.

The rest of the paper is organized as follows: In Section II, the problem is described in addition to a static game for jamming attack. Section III calculates Nash equilibrium for secondary and malicious users in both situations of complete and partial information about the rival. Meanwhile, numerical results are studied in section IV. Section V provides the conclusions and addresses the future works.

## II. SYSTEM MODEL

We consider a CRN with $n_N$ allocated spectrum bands. In this network, there are one secondary, one malicious and $n_p$ primary users. It is assumed that both malicious and secondary users know the number of primary users but they are unaware of their distribution in bands. Therefore, they suppose probability of occupation for each band by the primary user ($P_{pB}$) in each time slot, similar to a uniform distribution which is calculated in (1):

$$P_{pB} = \frac{n_p}{n_N} \qquad (1)$$

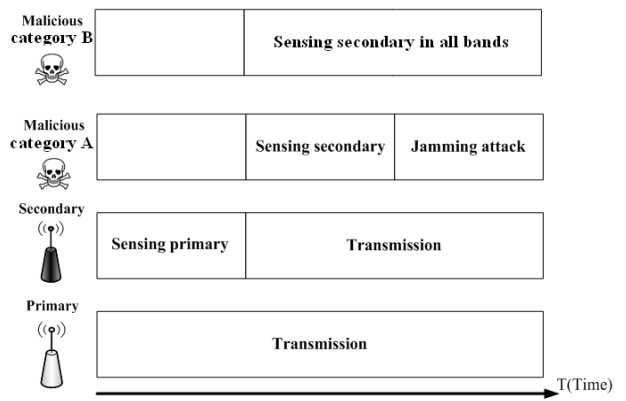

Fig. 1. The priority of spectrum band sensing for the primary, secondary and malicious users

In CRN, the primary users have higher priority over the secondary users to access to network. So the secondary users can't use the network as long as the primary users are present.

We proposed a discrete model to study CRN. In discrete-time model, time is divided into time slots and it assumed all users select their optimal action at the beginning of the time slots. Fig. 1 shows timing of the secondary, malicious and primary users to access to bands [6]. At the beginning of time slot, the secondary users sense the channel to discover whether the primary users are present or not. The secondary users can use the channel to send their information in absence of primary users.

However, at the beginning of time slot, the malicious user senses the band to distinguish presence or absence of the secondary user. Thus the malicious user waits at the beginning of each time slots where the secondary users start to send information. If the malicious user finds the secondary user, he/she will jam to the secondary user. Otherwise, the malicious user searches other bands to locate the secondary user. It is assumed that disrupting communication of the secondary user is of great importance for the malicious user and it has also the ability to sense all bands in one time slot. The malicious user doesn't need to sense the primary user because the task of the malicious user is to disturb communication of the secondary user.

The malicious user can jam to the secondary user to disrupt its communication. When they are both in the same band, a simple strategy for the secondary user to prevent a jamming attack would be switching to other bands. However, the malicious user can switch between different bands to find the secondary user. Thus, game theory is used among malicious and secondary users in a typical jamming attack to analyze the interactions.

Both secondary and malicious user may stay in their bands or switch to other bands. For game theoretical analysis, we should use gains and costs for both secondary and malicious users for their different actions [8]. $C_m$ and $C_s$ are the cost values of malicious and secondary users for switching to other bands, respectively. Gain value of the malicious user by performing a successful jamming attack is denoted by $G_m$. On the other hand, gain value of the secondary user to communicate in bands without either malicious or primary users is shown by $G_s$. Since to reestablish a communication has some costs for the secondary user, If a secondary user is jammed with the malicious user, it will lost $L_s$ According to availability of the secondary, malicious and primary users, there are different states for bands.

In some states, the secondary and malicious users don't have adequate information about the band of rivals while they are aware of this band in other states. Depending on the bands of rivals are known or not, these states are classified to three categories. It was demonstrated that analysis of the interaction on the malicious and secondary users is similar for different states of each category. Thereby, the three categories could be presented in addition to game theoretical analysis of the jamming attack in each category.

*A. Category A (Aware – Aware)*

In Category A, the band of rival is known for both secondary and malicious users. Only one of the six states has a situation similar to category A that is shown in fig. 2a.

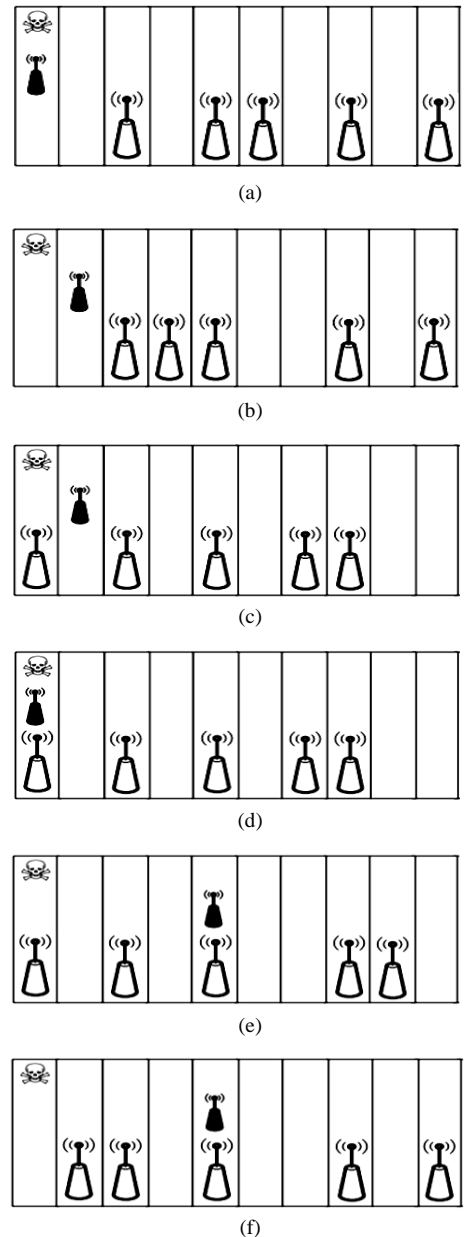

Fig. 2. Different states for the bands

This state, when the secondary user is jammed with malicious user and has lost $L_s$. In this situation, for next time slot, the secondary user should decide to stay in the band or try to find a vacant band to switch. Also the malicious user should opt to stay in the band or switch to another band for finding the secondary user in the next time slot. Thus both secondary and malicious users have two possible strategies in category A: staying in their band or switching to other band with uniform distribution. Based on fig.1, both secondary and malicious users should select their strategy at the beginning of slot time simultaneously, so their interaction could be modeled whit a static game.

To analysis of the jamming attack for category A, a static game is used. First the utility functions of malicious and secondary users are calculated which are summarized in table I. In table I, $P_{js}$ is the occupation probability of a band Just with the Secondary user in absence of both primary and malicious users, which is calculated by (2).

TABLE I. THE UTILITY FUNCTIONS FOR DIFFERENT STRATEGIES IN CATEGORY A

| Secondary \ Malicious | Switch to other bands | Stay in its band |
|---|---|---|
| Switch to other bands | $-C_s + (G_s \times P_{js}) - (L_s \times P_{jsm})$ , $-C_m + (G_m \times P_{jsm})$ | $-C_s + G_s \times (1 - P_{pB}), 0$ |
| Stay in its band | $G_s \times (1 - P_{pB}), -C_m$ | $-L_s \times (1 - P_{pB}), G_m \times (1 - P_{pB})$ |

TABLE II. THE UTILITY FUNCTIONS FOR DIFFERENT STRATEGIES IN CATEGORY B

| Secondary \ Malicious | Stay in its band | Switch to secondary's band |
|---|---|---|
| Switch to other bands | $-C_s + (G_s \times P_{js}) - (L_s \times P_{jsm})$ , $G_m \times P_{jsm}$ | $G_s \times (1 - P_{pB}), -C_m$ |
| Stay in its band | $G_s \times (1 - P_{pB}), 0$ | $-L_s + (1 - P_{PB}), G_m \times (1 - P_{pB}) - C_m$ |

$$P_{js} = 1 - (\frac{1}{n_N - 1} + P_{pB} - \frac{P_{pB}}{n_N - 1}) \qquad (2)$$

In (2), $\frac{1}{n_N - 1}$ is the occupation probability of other bands with the malicious user when he/she switches to other bands. Meanwhile $P_{jsm}$, in table I is the occupation probability of other bands **J**ust with **S**econdary and **M**alicious users in absence of the primary user, which is given by (3).

$$P_{jsm} = \frac{1}{n_N - 1} \times (1 - P_{pB}) \qquad (3)$$

*B. Category B (Aware – Unaware)*

In category B, the band of secondary user is known for the malicious user whereas the band of malicious user in unknown for the secondary user and we noted it with (Aware – Unaware). Two states out of six states have the situation of category B and are showed in fig. 2b and fig. 2c. The malicious user in category B first searches for the secondary user in its band and doesn't find it, so the malicious user should sense other bands to find the band of secondary user.

The cause of sensing other bands is that the malicious user wants to identify the band of secondary user and increase his/her chance for a successful jamming in the next time slot. However the secondary user doesn't know the band of malicious user, he/she is aware that the malicious user knows his/her band. The secondary user doesn't sense other bands for discovering the band of malicious user because the secondary user is aware that the malicious user doesn't send jamming signal in situations of category B.

The secondary user has two choices in category B: switch to other bands with uniform distribution; or stay in his/her current band. Although the malicious user has two strategies: staying in his/her band; or switching to the band of secondary user.

The malicious user doesn't have any other choice for switching to other bands (except of his/her band and the band of secondary user), because the malicious user knows that the secondary user will switches to another band with uniform distribution, so the malicious user has equal chance to jam the secondary user in these bands. Analysis of attack for category B can be model by a static game. The utility functions are listed in table II.

*C. Category C (Unaware – Unaware)*

In category C, both secondary and malicious users are unfamiliar with the band of rival and we noted it with (Unaware – Unaware). Three states out of six states are placed in category C, which are shown in fig. 2d, fig. 2e and fig. 2f.

In various situations of category C, the malicious user senses his/her band as well as other bands to find the band of secondary user, while the malicious user doesn't find the band of secondary user because the secondary user doesn't transmit any information toward the channel. Thus, the malicious user knows that secondary and primary users are located in the same band. Meanwhile, the secondary user is unaware about the band of malicious user.

Secondary and malicious users don't have any information about the band of rival in category C, so it is rational for them to stay in their band and not to waste their energy for switching to other bands. As a result, the optimal strategy for secondary and malicious users would be to stay in their band for the next time slot. Since the primary users occupy the bands having uniform distribution, the network can be changed to category A and B for the next time slot. The state of the network is one of 6 states or one of 3 categories. For next time slot, state of the network can change to another categories or stay without change. Fig.3 depicts these three categories with transitions of the network among them.

Obviously, interaction between malicious and secondary users is based on complete/partial information about the rival. So, one must analyze the jamming attack in both complete and partial information.

III. THE PROPOSED GAME

First, we consider a situation with complete information about rival's cost and gain. So, we should analyze a simple static game. Then we analyze an unknown jamming attack and show that two Nash equilibriums are equal. The method of calculating Nash equilibrium is similar for these two categories, so we used from *a, b, c, d, e, f, g* and *h* (in table III) instead of utility functions to solve the Nash Equilibrium for both categories A and B. For example $a$ is equal $-C_s + (G_s \times P_{js}) - (L_s \times P_{jsm})$ in category A (from table I) and is equal $-C_s + (G_s \times P_{js}) - (L_s \times P_{jsm})$ in category B (from table II).

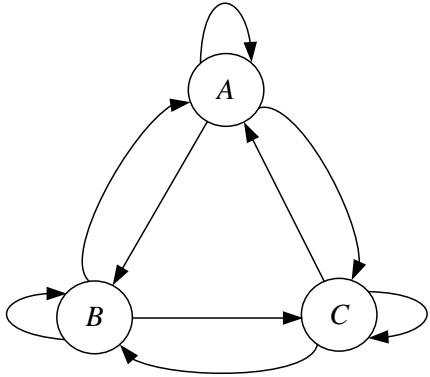

Fig. 3. Different categories for the network Nash equilibrium

For solving this simple static game (table III), it is assumed that the secondary user selects his/her first strategy with probability $p$ while the malicious user chooses his/her first strategy with probability $q$. If the secondary user selects his/her first strategy, then it would have a utility value of $U_{s1}$. So, if the secondary user chooses his/her second strategy, then it has a utility value equal to $U_{s2}$. $U_{s1}$ and $U_{s2}$ are calculated in (4) and (5), respectively.

$$U_{s_1} = a \times q + b \times (1-q) \quad (4)$$

$$U_{s_2} = c \times q + d \times (1-q) \quad (5)$$

Similarly, $U_{m1}$ and $U_{m2}$ are utilities of the malicious user, when choosing his/her first and second strategies, respectively.

$$U_{m_1} = e \times p + g \times (1-p) \quad (6)$$

$$U_{m_2} = f \times p + h \times (1-p) \quad (7)$$

Nash equilibrium for a static game occurs when utilities of the first strategy and the second strategy are equal. Thus:

$$U_{s_1} = U_{s_2} \Rightarrow q = \frac{d-b}{a-c+d-b} \quad (8)$$

$$U_{m_1} = U_{m_2} \Rightarrow p = \frac{h-g}{e-f+h-g} \quad (9)$$

TABLE III. THE UTILITY FUNCTIONS FOR DIFFERENT STRATEGIES

| Malicious / Secondary | First strategy | Second strategy |
|---|---|---|
| First strategy | $a, e$ | $b, f$ |
| Second strategy | $c, g$ | $d, h$ |

Secondary and malicious user may have no information about cost of the rival and gain values. For example, when the new malicious user enters the network, both secondary and malicious users would have partial information about his/her rival.

In this situation (partial information), we can use learning games like fictitious play to calculate Nash equilibrium. Fictitious play is a historical-based game in which both malicious and secondary users should save some information about their rival.

For this reason, it is supposed that $h_{s1}$ and $h_{s2}$ are the number of times that the secondary user chooses his/her first and second strategies, respectively and the malicious user should count $h_{s1}$ and $h_{s2}$. Meanwhile, $h_{m1}$ and $h_{m2}$ are the number of times that the malicious user selects his/her first and second strategies, respectively and the secondary user should count $h_{m1}$ and $h_{m2}$.

At the beginning of each stage in this game, the secondary user calculates the expected utility for choosing his/her first $U_{s1}^{ex}$) and second ($U_{s2}^{ex}$) strategies like (10) and (11), respectively [11].

$$U_{s_1}^{ex} = (a \times h_{m_1}) + (b \times h_{m_2}) \quad (10)$$

$$U_{s_2}^{ex} = (c \times h_{m_1}) + (d \times h_{m_2}) \quad (11)$$

If $U_{s_1}^{ex} > U_{s2}^{ex}$, then the secondary user should choose the first strategy. Also, when $U_{s_1}^{ex} < U_{s2}^{ex}$, the secondary user should adopt the second strategy. Thus, when $U_{s_1}^{ex} = U_{s2}^{ex}$, the secondary user should randomly select between first and second strategies.

Moreover, the malicious user calculates the expected utility for choosing his/her first ($U_{m1}^{ex}$) and second ($U_{m2}^{ex}$) strategies based on (12) and (13), respectively.

$$U_{m_1}^{ex} = (e \times h_{s_1}) + (g \times h_{s_2}) \quad (12)$$

$$U_{m_2}^{ex} = (f \times h_{s_1}) + (h \times h_{s_2}) \quad (13)$$

Similarly, when $U_{m_1}^{ex} > U_{m_2}^{ex}$, the malicious user should pick the first strategy. If $U_{m_1}^{ex} < U_{m_2}^{ex}$, the malicious user should select the second strategy and when $U_{m_1}^{ex} = U_{m_2}^{ex}$, the -malicious user should follow his/her first and second choices randomly. With this algorithm, $p^*$ and $q^*$ could be calculated from (14) and (15) for partial information situation, respectively.

$$p^* = \frac{h_{s_1}}{h_{s_1} + h_{s_2}} \quad (14)$$

$$q^* = \frac{h_{m_1}}{h_{m_1} + h_{m_2}} \quad (15)$$

It can be proved that by increasing the number of repetitions for a category, $p^*$ and $q^*$ for the partial information situation will converge to $p$ and $q$ for complete information situation, respectively [11]. This convergence is also shown in the numerical results.

Fig. 3 depicts change in the status of network for three categories. So, with the previously mentioned algorithm for fictitious play, the secondary (malicious) user in each of A and B categories, observes and saves actions of the malicious (secondary) user. It is important to notice that each of the categories A and B have their own history about the action of rival. For example, if they are located in one of the two states of category B (fig. 2b and fig. 2c), they should save the action of rival in the history of category B. Obviously they don't save the action of rival for category C because their optimal strategy is to stay in their bands there. Also when they are in state of category A, its history will be complete.

The observations of secondary and malicious user about the action of rival are rather partial. If the state of network is A and in the next time slot it is A again, both secondary and malicious users will be aware about actions of each other. Thus they keep actions in the history of state A. If state of network is A and in the next time slot it is B, actions of secondary and malicious users will be known for both of them. Thus they keep actions in the history of state B. When the state of network is changed from state A to B or remains in state B in the coming slot time, the malicious user observes action of the secondary user and updates his/her history of states A and B.

For the abovementioned situation, the secondary user can view actions of the malicious user if stays in its own state for the next time slot. If the secondary user switches to other bands in the next time slot, it cannot view actions of the malicious user. Therefore, histories of the secondary user are usually rather smaller than those of the malicious user. This causes the secondary user to reach Nash equilibrium later than the secondary user.

In other states, secondary and malicious users can't observe the actions of each other and they can't either update their history.

However, when the network is in state C or switches from state C to other states, they don't need to record their actions of rival because they know their optimal strategy in state C.

## IV. NUMERICAL RESULTS

This section will discuss the numerical results of jamming attacks. Initial values for various parameters used in this analysis were: $n_N = 10$, $n_p = 5$, $C_s = 5$, $C_m = 2$, $G_s = 50$, $G_m = 75$ and $L_s = 100$.

According to the initial values used before as well as (8) and (9), the Nash equilibrium for categories A and B are $p = 0.94$ and $q = 0.84$, and $p = 0.85$ and $q = 0.85$, respectively.

Fig. 4 and fig. 5 show the convergence of fictitious play to Nash equilibrium for secondary and malicious users in both categories A and B. For secondary and malicious users, the strategy of fictitious play is different from Nash equilibrium in the early stages. But by increasing stages of game and updating their histories, their strategies will approach much closer to Nash equilibrium.

## V. CONCLUSION

The CR network is introduced to solve the problem of small spectrum band for data transmission. Jamming attack is one of the common security problems in CR networks. In this paper, a game theoretical model is proposed to analyze the jamming attack.

In the proposed model, it has been shown how secondary and malicious users can obtain the maximum utility simultaneously. Meanwhile, behaviors of secondary and malicious users were analyzed, when there is partial information about rival.

The results show that by increasing the history of attacks, the equilibrium of game in converged to the Nash equilibrium, thus this game is suitable when the information of attack is partial.

Malicious users can join together to conduct a cooperative attack against CR. Analysis of such a cooperative attack is recommended for future works.

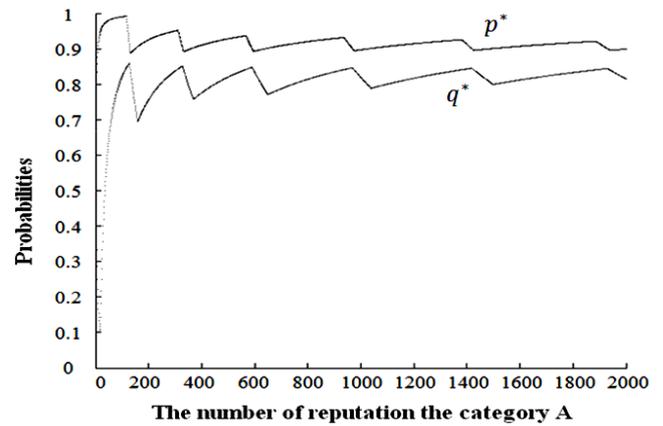

Fig. 4. Convergence of the secondary and malicious user's strategies in category A

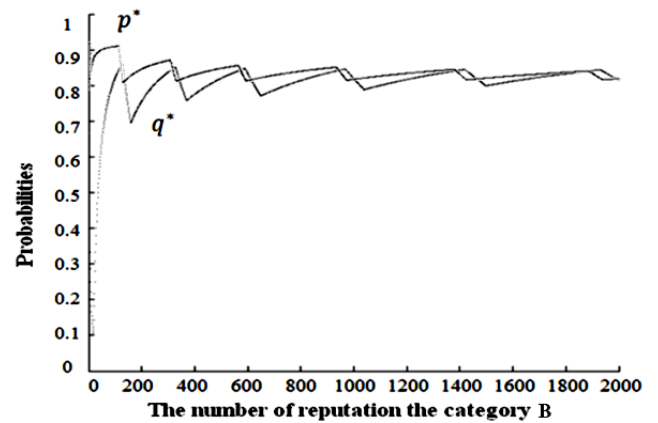

Fig. 5. Convergence of the secondary and malicious user's strategies in category B